\documentclass[a4paper]{article}
\pdfoutput=1

\usepackage{float}       
\usepackage{helvet}      
\usepackage{mathptmx}    
\usepackage{rsc}         
\usepackage{setspace}    
\AtBeginDocument{\doublespacing}

\newfloat{chart}{htbp}{loc}
\newfloat{graph}{htbp}{loh}
\newfloat{scheme}{htbp}{los}

\usepackage{graphicx}

\author{
\centerline{Ibrahim Abou Hamad$^{1,}$\footnote{Current address: Department of 
Physics and Center for Materials Research and Technology, Florida State 
University, Tallahassee, FL 32306-4350, USA, {\it E-mail address:\/} iabouhamad@fsu.edu}, 
M. A. Novotny$^{1,2}$, D. Wipf$^3$, and P. A. Rikvold$^4$}\\
\centerline{
$^1$HPC$^2$, Center for Computational Sciences, Mississippi State University,
Mississippi State, Mississippi 39762}\\
\centerline{$^2$Department of Physics \& 
Astronomy, Mississippi State University, Mississippi State, Mississippi 39762}\\
\centerline{$^3$Department of Chemistry, Mississippi State University, 
Mississippi State, Mississippi 39762}\\
\centerline{$^4$Department of Physics and Center for Materials Research and 
Technology}\\
\centerline{Florida State University, Tallahassee, Florida 32306-4350}}

\title{A new battery-charging method suggested by molecular dynamics
  simulations}

\begin{document}

\maketitle

\begin{abstract}
  Based on large-scale molecular dynamics simulations, we propose a new
  charging method that should be capable of charging a Lithium-ion
  battery in a fraction of the time needed when using traditional
  methods. This charging method uses an additional applied oscillatory
  electric field. Our simulation results show that this charging method
  offers a great reduction in the average intercalation time for
  Li$^+$ ions, which dominates the charging time. The oscillating
  field not only increases the diffusion rate of Li$^+$ ions in the
  electrolyte but, more importantly, also enhances intercalation by
  lowering the corresponding overall energy barrier.
\end{abstract}

\section{Introduction}

The widespread use of Lithium batteries with liquid electrolytes in portable
electronics, as well as their potential for use in environmentally friendly
electric vehicles, make Lithium-battery technology a field of active
investigation~\cite{Armand2008,Xu2004}. At the same time, revolutionary
developments in computer hardware and algorithms now enable
simulations with millions of individual particles~\cite{vashishta2006}. Computer simulations, ranging from
molecular dynamics (MD) to Monte Carlo (MC) simulations and quantum-mechanical
density functional theory are therefore becoming widely used to model
electrochemical systems~\cite{Perbook} and guide experimental research.
Quantum-mechanical simulations have been used to study the average
voltage needed, the charge transfer~\cite{Aydinol1997}, and the phase 
diagrams~\cite{Ceder1999} for Lithium intercalation in transition metal oxides. Studies of a
graphite electrode with intercalated Lithium~\cite{Marquez1998} and of surface chemistry
at the electrode-electrolyte interface~\cite{Wang2002} have also used quantum-mechanical
simulations. MD and
MC~\cite{Darling1999} simulations have contributed important insight about
the interfacial structure, structural changes and Lithium ion diffusion
in Lithium batteries in the anode~\cite{Marquez2007,Marquez2001} or cathode~\cite{Garcia1998,Garcia1999} half cell. These studies were all performed under
stationary voltage conditions.

The Lithium-ion battery-charging process is marked by the
intercalation of Lithium ions into the anode material. Here we present
large-scale molecular dynamics simulations of this process under \textit{oscillating} voltage conditions. These
simulations suggest a new charging method that has the potential to
deliver much shorter charging times, as well as the possibility of
providing higher power densities. Moreover, it is argued that while
the chemical nature of the electrodes decides the energy output, the
electrolyte, in most situations, controls the rate of mass flow of the
ions and thus how fast energy can be stored or
released~\cite{Linden2001}. Our simulations suggest that while faster
diffusion contributes to faster charging, the rate-limiting step is
the Li-ion intercalation into the graphite anode.

The rest of this paper is organized as follows. The details of the
molecular-dynamics simulations, as well as the composition of the
model system are presented in section~\ref{MD}. The simulation results
and a discussion of diffusion and intercalation times are presented in
section~\ref{results}. The paper ends with a conclusions section. 

\section{Model and simulation method}
\label{MD}

The model system for the anode half-cell of a Lithium-ion battery 
is composed of an anode represented by a stack of graphite sheets, an 
electrolyte of ethylene carbonate and propylene carbonate molecules, and 
Lithium and  hexafluorophosphate ions. After reaching a constant volume with 
simulations in the NPT ensemble, long production runs are done in the NVT
ensemble at room temperature.

\subsection{Molecular Dynamics}

Molecular Dynamics is based on the solution of the classical equations 
of motion for a system of $N$ interacting atoms 
\cite{Allen:92,Haile:92,Rapaport:04}. From the potential energy $E_{\rm P}$,
the instantaneous force, acceleration, and velocity for the $i$th atom are 
calculated, and its position is updated accordingly. This process is repeated
for all atoms at each time step.

The General Amber Force Field (GAFF)~\cite{GAFF:04} was used
to approximate the bonded and van der Waals' interactions of all the simulation 
atoms, while
the simulation package Spartan (Wave-function, Inc., Irvine, CA) was used at
the Hartree-Fock/6-31g* level to obtain the necessary point charges for the
ethylene carbonate, propylene carbonate and PF$_{6}^{-}$. The charges were
sligtly modified to make the charges equal on identical atoms while keeping
the sum of the charges constant. The AMBER program {\it tleap\/} was used to 
build the molecules and set their force field parameters and combine them 
into one parameter input file for the system as a whole. The 
simulation package NAMD~\cite{NAMD} was used for the MD simulations, while
VMD~\cite{VMD} was used for visualization and analysis of the system. NAMD
was primarily chosen because it is a parallel MD program that scales
remarkably well with the number of processors. 

\subsection{Model System}

\begin{figure}
\vspace{-0.01truecm}
\begin{center}
\includegraphics[angle=0,height=.45\textwidth]{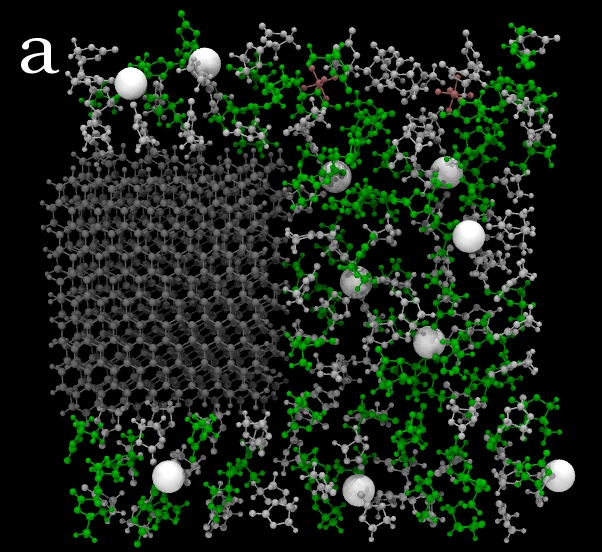}
\hspace{0.01truecm}
\includegraphics[angle=0,height=.45\textwidth]{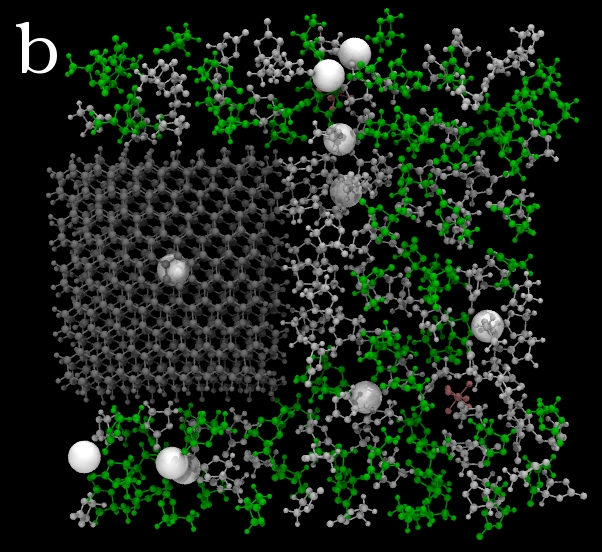}
\vspace{0.1truecm}
\end{center}
\caption[]{\small
The system was first simulated in the NPT ensemble 
until it reached constant volume. The system has periodic boundary 
conditions and is simulated at one atm and 300 K. Top view, 
perpendicular to the plane of the graphite sheets
{\bf (a)} Snapshot after $200$ ns MD simulation in the NVT ensemble of the model
system containing four graphite sheets, two PF$_{6}^{-}$ ions and ten Li$^{+}$ 
ions, solvated in $69$ propylene carbonate and $87$ ethylene carbonate
molecules. {\bf (b)} Snapshot after 19 ns MD simulation with an additional 
oscillating electric field. 
}
\label{fig:MDsystem}
\end{figure}

The anode half-cell is modeled using four graphite
sheets containing $160$ carbon atoms each (anode), two PF$_{6}^{-}$ ions, and
ten Li$^{+}$ ions, solvated in an electrolyte composed of $69$ propylene
carbonate
and $87$ ethylene carbonate molecules (see Fig.~\ref{fig:MDsystem}). The
anode sheets were fixed from the side by keeping the positions of the carbon 
atoms at one edge of each sheet fixed. The charges on the carbon atoms of the 
graphite sheets was set to $-0.0125$ \textit{e} per atom to simulate a charging 
field.

The simulations proceeded as follows. The system energy was minimized, after
which the simulations were run at constant pressure (in the NPT ensemble) using
a Langevin piston Nos\'{e}-Hoover method~\cite{Martyna:94,Feller:95} as
implemented in the NAMD software package until the system had reached its
equilibrium volume at a pressure of $1$ atm and 300 K. The system's 
behavior was then simulated for $200$ ns ($100$ million steps)
in the NVT ensemble. A $2$ fs timestep was used. The electrostatic interactions 
were calculated by the particle-mesh Ewald method~\cite{Darden93,Essmann95}.  

The Li$^{+}$ ions stay
randomly distributed within the electrolyte, and none of the Li$^{+}$
ions intercalates between the graphite sheets after $200$ ns
(see Fig.~\ref{fig:MDsystem}{\bf(a)}). This behavior was confirmed by $10$ independent runs.
While it is expected that, given sufficient time, the 
Li$^{+}$ ions would move closer to the electrode and intercalate between
the graphite sheets, they do not show such behavior on the time scales currently
accessible by molecular dynamics simulations.

For intercalation to occur, the Lithium ion has first to diffuse within
the electrolyte until it reaches the graphite electrode, and second to overcome
the energy barrier at the electrode-electrolyte interface. 
In order to facilitate intercalation, a new charging
method was explored.  An external oscillating square-wave field (amplitude
$A =$ 5 kCal/mol, frequency $f =$ 25 GHz) was applied in the direction 
perpendicular to the plane of the graphite sheets. This oscillating field is
in addition to the charging 
field due to the fixed charge on the graphite carbons. This additional field not
only increases diffusion, but also causes some of the 
Lithium ions to intercalate into the graphite sheets within an average time of
about 50 ns for the amplitude and frequency mentioned above, calculated 
from 100 independent runs.

\section{Simulation Results}
\label{results}

The simulations were performed on 2.6 GHz Opteron processors, using 160 
processors at a time. The amplitude of the oscillating field was fixed
at values from $0.9 A$ to $1.3 A$ in steps of $0.1$. For each value of the amplitude
ten simulations were collected. For the value of the amplitude equal to $A$,
$100$ different runs were simulated. The total computer time required was
approximately 4$\times 10^5$ CPU hours. The data were used to study the 
dependence of the diffusion and intercalation times on the field amplitude.
A study of this dependence on the field frequency is currently initiated~\cite{Hamad2010}.

\subsection{Diffusion}

\begin{figure}
\vspace{-0.01truecm}
\begin{center}
\includegraphics[angle=0,height=.45\textwidth]{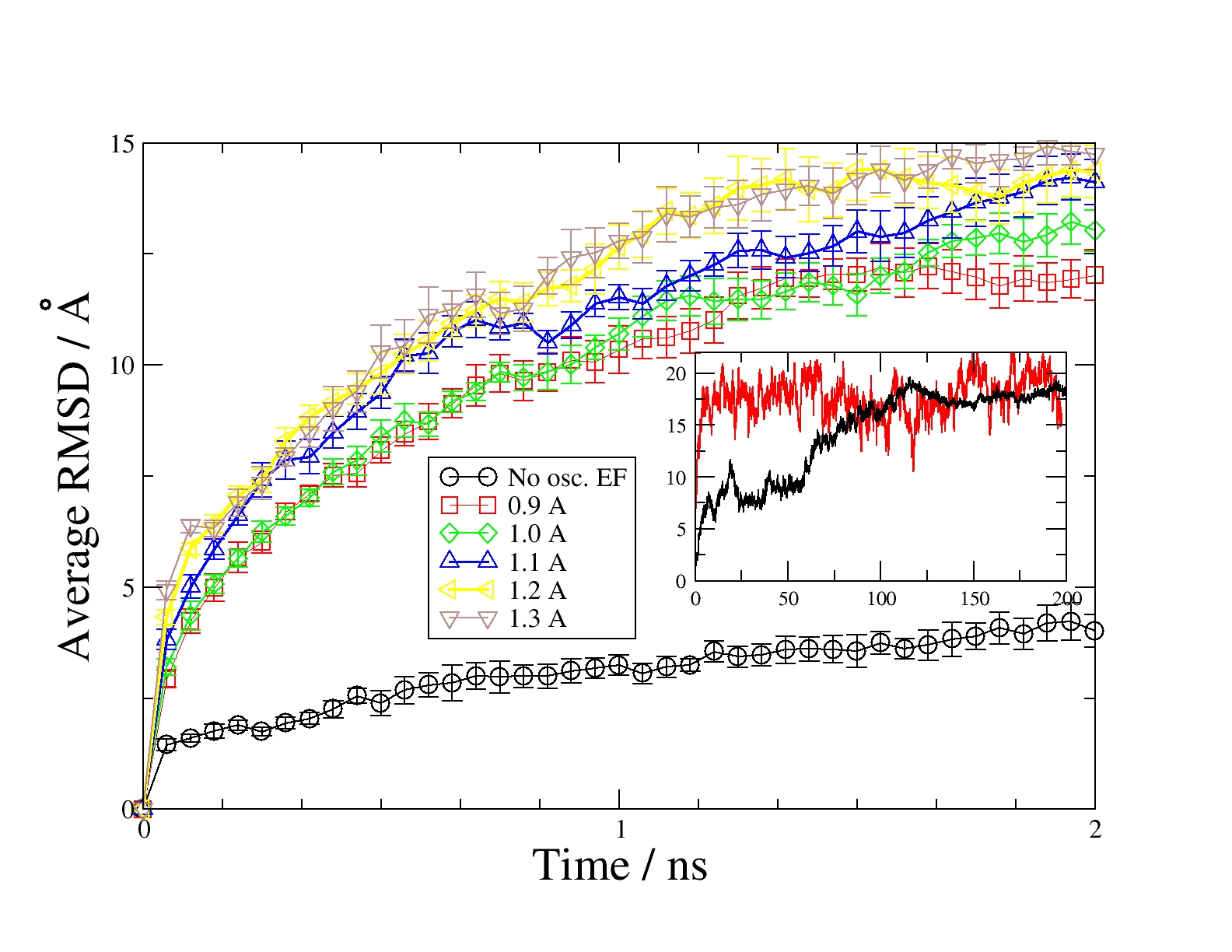}
\vspace{0.1truecm}
\end{center}
\caption[]{\small
Average root-mean-square displacement of Lithium ions as a function of time.
Diffusion is much faster with the additional oscillating field with frequency
25 GHz for all values
of its amplitude ($\times A$ with $A = 5$ kCal/mol). A comparison between
two long simulation runs with and without the oscillating field is shown in
the inset. Notice that the lower curve (without oscillating field) only
approaches the upper curve (with oscillating field) after about $100$ ns of
simulation time.
}
\label{fig:RMSD}
\end{figure}

The use of the additional oscillating field leads to much faster diffusion. 
Figure~\ref{fig:RMSD} shows the average root-mean-square displacement of the 
Li$^{+}$ ions as a function of time. The average is over all the ten Li$^{+}$ ions
and over the number of different simulations for each value of the amplitude of the
electric field. When using an oscillating electric field, within the first $2$ ns of simulation the root-mean-square displacement reaches
a value close to its limiting value defined by the finite size of the system, while much slower diffusion is seen
for the simulations without that additional field.
The inset shows a comparison of the root-mean-square displacement for a single 
run with and without the oscillating electric field of frequency 25 GHz and 
amplitude $A$ for a much longer
time period. The root-mean-square displacement for simulations without the
oscillating electric field start approaching the limiting value only after about
$100$ ns of simulation time. The diffusion coefficient estimated from the
root-mean-square displacement of Li ions in the electrolyte, from simulations
without an oscillating electric field, is about $4x10^{-12}$ m$^2$/s, which is 
consistant with the results from a different MD study by~\citeauthor{Marquez2007}.~\cite{Marquez2007}
 
\subsection{Intercalation time}

\begin{figure}
\vspace{-0.01truecm}
\begin{center}
\includegraphics[angle=0,height=.75\textwidth]{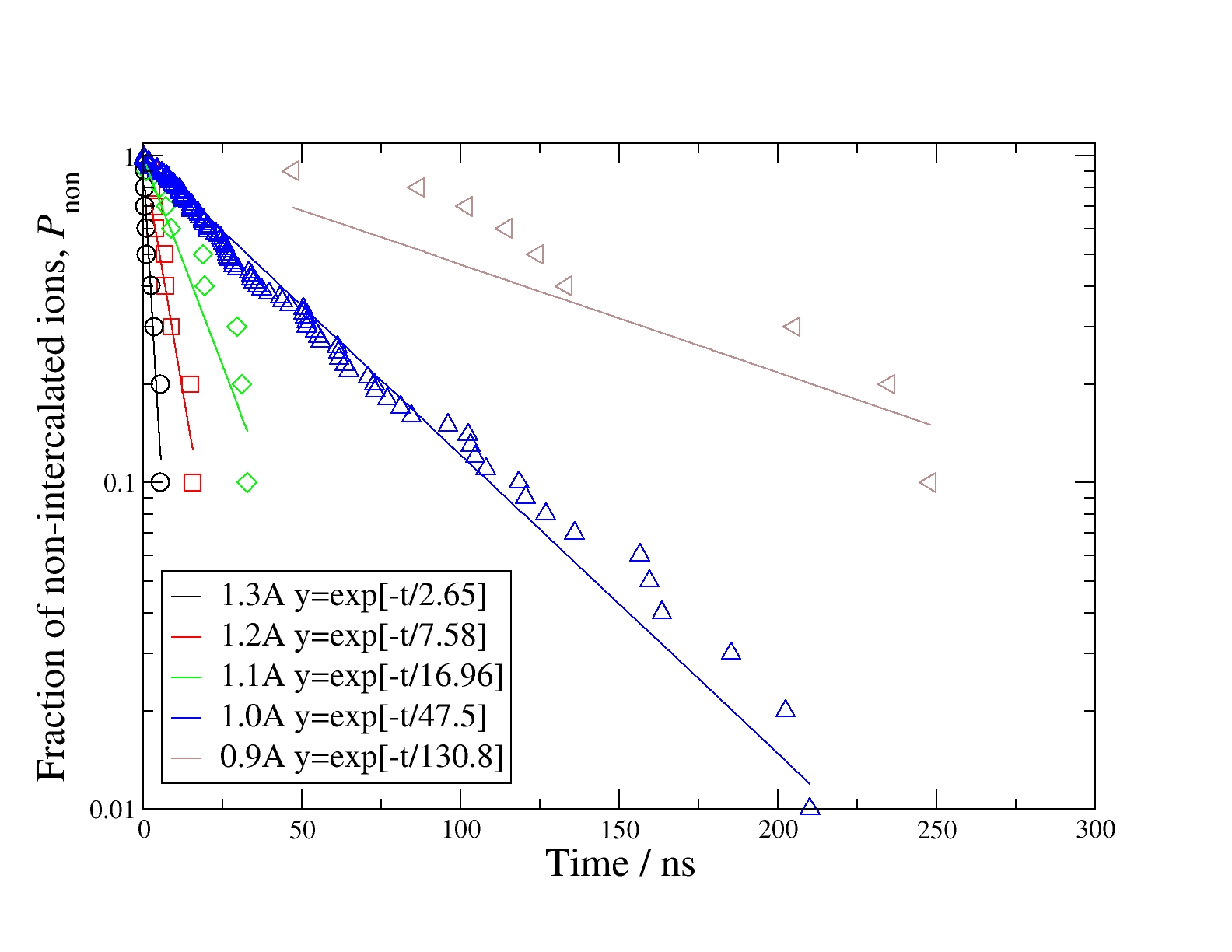}
\vspace{0.1truecm}
\end{center}
\caption[]{\small
Fraction of non-intercalated Lithium ions versus time. The lines are fits to
the 
form given by Eq.~(\ref{eq:poison}), and the fitted parameters are shown in
the legend for the different amplitudes of the applied oscillating field. The
denominator in the exponential is the average intercalation time in ns.
}
\label{fig:intercalation}
\end{figure}

\begin{figure}
\vspace{-0.01truecm}
\begin{center}
\includegraphics[angle=0,width=8.3 cm]{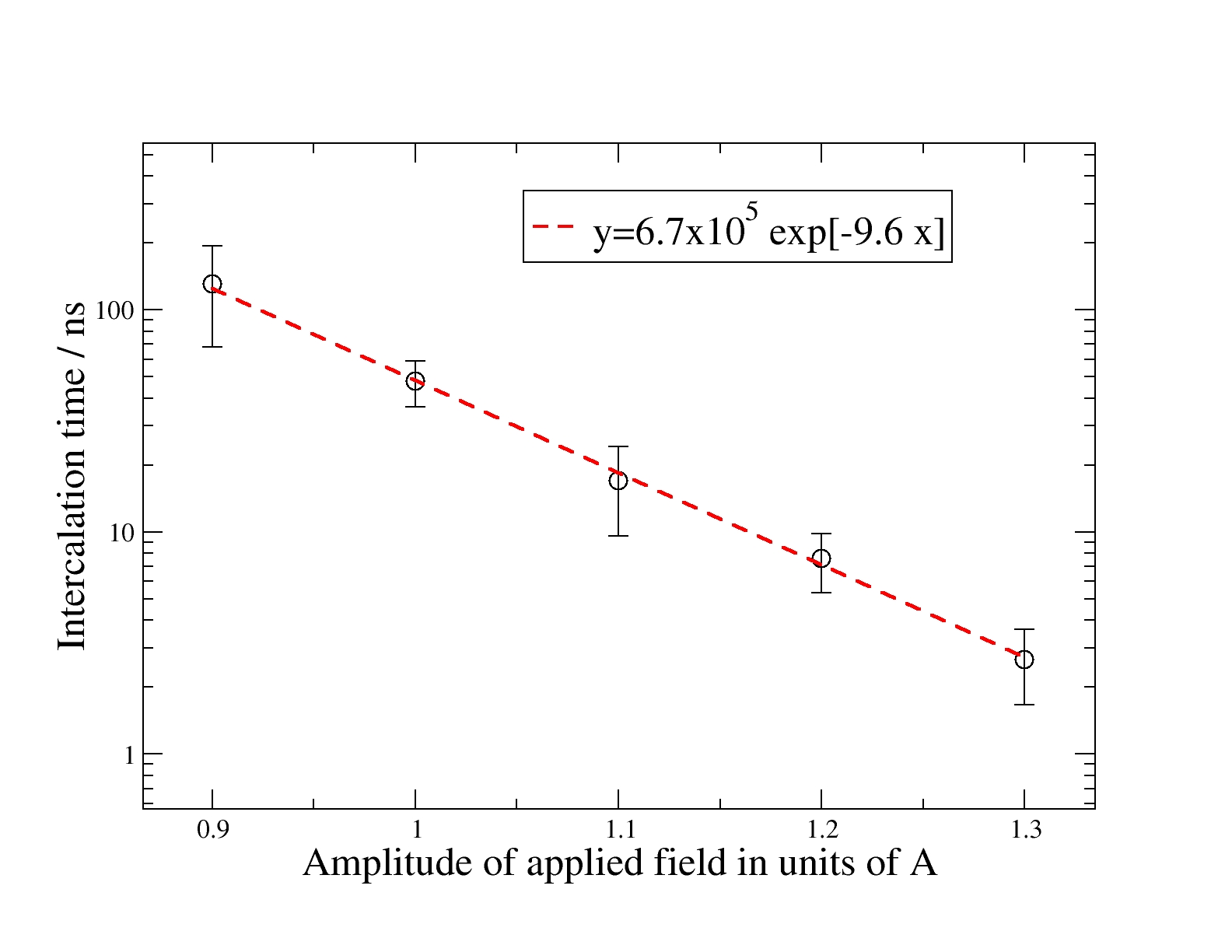}
\vspace{0.1truecm}
\end{center}
\caption[]{\small
The exponential dependence of the average intercalation
time on the amplitude of the applied field. The dashed line is a fit to the
exponential form with parameters given in the legend.
}
\label{fig:Arh}
\end{figure}

The inclusion of an additional oscillating electric field not only increases
the diffusion rate, but also leads to the intercalation of Lithium between the
graphite sheets. The intercalation events of the Lithium ions are considered
to be a Poisson process~\cite{cox:65}. Then the fraction of the number of
runs simulated, in which \textit{no} ions have intercalated by time $t$, is given by
\begin{equation}
P_{\rm non} = e^{(-t/\tau)},
\label{eq:poison}
\end{equation}
where $\tau$ is the average intercalation time for that process. Thus, the
average intercalation time can be obtained from a fit of the fraction of
non-intercalated ions as a function of time to the form given by
equation~\ref{eq:poison}.
A plot of the fraction of non-intercalated ions versus time, as well
as the fitting lines and average intercalation times are shown in
Fig.~\ref{fig:intercalation}.
The frequency used was always 25 GHz. The standard
deviation is assumed to be on the order of $y$, and the error bars on the
fitting paramater are calculated using a $1/y^2$ weighted linear-regression 
analysis (Fig.~\ref{fig:Arh}).

The dependence of the average intercalation time on the amplitude of the applied
oscillating electric field is shown in Fig.~\ref{fig:Arh}. The average intercalation
time shows an exponential dependence on the amplitude of the
applied field. This dependence is consistent with a linear decrease of
the effective free-energy barrier against intercalation with increasing 
amplitude of the applied field.
However,  due to the small
window of amplitude values, other forms of the dependence of the barrier on the field amplitude cannot be excluded. Moreover, care must be 
taken in extrapolating this dependence outside of the simulated amplitude range.
Extrapolating to a large amplitude of about $2.78$ $A$,
the time would be equal to the MD time step. In the more interesting
regime of small amplitudes, extrapolating to zero amplitude would lead to an
average intercalation time of $6.7\times10^5$ ns. This time scale is clearly larger
than the maximum time for which MD simulations of a system of this complexity can
presently be performed. For no oscillatory field, as in Fig.~\ref{fig:MDsystem}{\bf (a)},
we performed simulations for a total time of about 200 ns without observing any intercalations. These simulations were performed with a uniform charge
distribution on the C atoms. Since graphite is a conductor, it is likely that
the charge distribution is biased toward the electrode surface. To check the
influence of the charge distribution, we therefore performed exploratory
simulations with the electronic charge distributed over just the two C layers
near the interface. In these simulations (Amplitude 1.3 $A$) we found only
an insignificant change in the diffusion time, while the intercalation time
increased by approximately a factor of three. However, this is still very much
shorter than the constant-field case. Thus, the qualitative effect of the
oscillating field is independent of the details of the charge distribution.

The charging time in our simulations depends exponentially on 
the amplitude of the oscillating field, while diffusion did not differ much for
different values of the amplitude (see Fig.~\ref{fig:RMSD}). Included in the
intercalation times are the time to diffuse toward the electrode and the time to
overcome the electrode-electrolyte interface barrier. This leads us to two
conclusions. First, the oscillating field not only increases the diffusion 
rate, but it also lowers the free-energy barrier for intercalation.  Second, 
while diffusion is important for faster charging, the rate-limiting 
step for our case is the intercalation process. 

\section{Conclusions}

In this paper we have proposed a new charging method for Lithium-ion batteries
that uses an additional oscillating electric field to reduce the average intercalation time, and thus the charging time. The dependence of the intercalation time on
the applied field amplitude is exponential, and thus there is the potential for very fast charging times.

This exponential dependence indicates that while the oscillating field does 
increase the diffusion rate of Li$^+$ ions (not exponentially), more 
importantly it enhances intercalation through lowering the intercalation 
free-energy barrier. While faster diffusion is important for faster charging,
lowering the intercalation barrier at the electrode-electrolyte interface is 
seen to be more important in the frequency and amplitude regimes we were able
to simulate using molecular dynamics simulations. Future work will consider the
effect of different frequencies~\cite{Hamad2010}, as well as quantitative
comparison with electrochemical experiments where possible.

\section*{Acknowledgements}
This work was supported by U.S. National Science Foundation Grant No.
DMR-0802288 (Florida State University) and by the HPC$^2$ Center for 
Computational Sciences (Mississippi State University).
%
%
%
\bibliographystyle{rsc}
\bibliography{paper}

\end{document}